\documentclass[aps,prl,twocolumn,showpacs,superscriptaddress,groupaddress,amsmath,amssymb]{revtex4-1}

\usepackage{graphicx}  
\usepackage{dcolumn}   
\usepackage{bm}        
\usepackage{verbatim}   
\usepackage{textcomp}

\usepackage{xcolor} 

\begin{document}


\title{Experimental Constraint on Stellar Electron-Capture Rates from the ${}^{88}\text{Sr}(t,{}^{3}\text{He}+\gamma){}^{88}\text{Rb}$ reaction at 115~MeV/u}


\author{J. C.~Zamora}
  \affiliation{National Superconducting Cyclotron Laboratory, Michigan State University, East Lansing, MI 48824,
  USA}
\affiliation{Joint Institute for Nuclear Astrophysics: CEE,  Michigan State University, East Lansing, MI 48824,
USA}

\author{R.G.T.~Zegers}
\affiliation{National Superconducting Cyclotron Laboratory, Michigan State University, East Lansing, MI 48824,
  USA}
\affiliation{Joint Institute for Nuclear Astrophysics: CEE,  Michigan State University, East Lansing, MI 48824,
USA}
\affiliation{Departament of Physics and Astronomy, Michigan State University, East Lansing, MI 48824, USA}

\author{Sam M.~Austin}
\affiliation{National Superconducting Cyclotron Laboratory, Michigan State University, East Lansing, MI 48824,
  USA}
\affiliation{Joint Institute for Nuclear Astrophysics: CEE,  Michigan State University, East Lansing, MI 48824,
USA}

\author{D.~Bazin}
\affiliation{National Superconducting Cyclotron Laboratory, Michigan State University, East Lansing, MI 48824,
  USA}

\author{B. A.~Brown}
\affiliation{National Superconducting Cyclotron Laboratory, Michigan State University, East Lansing, MI 48824,
  USA}
\affiliation{Joint Institute for Nuclear Astrophysics: CEE,  Michigan State University, East Lansing, MI 48824,
USA}
\affiliation{Departament of Physics and Astronomy, Michigan State University, East Lansing, MI 48824, USA}

\author{P.C.~Bender}
\affiliation{National Superconducting Cyclotron Laboratory, Michigan State University, East Lansing, MI 48824,
  USA}

\author{H.L.~Crawford}
\affiliation{Lawrence Berkeley National Laboratory, Berkeley, CA 94720, USA}

\author{J.~Engel}
\affiliation{Department of Physics and Astronomy, The University of North Carolina at Chapel Hill, Chapel Hill, NC 27599, USA}

\author{A.~Falduto}
\affiliation{Departament of Physics, Central Michigan University, Mt. Pleasant, MI 48859, USA}
\affiliation{Joint Institute for Nuclear Astrophysics: CEE,  Michigan State University, East Lansing, MI 48824,
USA}

\author{A.~Gade}
\affiliation{National Superconducting Cyclotron Laboratory, Michigan State University, East Lansing, MI 48824,
  USA}
\affiliation{Joint Institute for Nuclear Astrophysics: CEE,  Michigan State University, East Lansing, MI 48824,
USA}
\affiliation{Departament of Physics and Astronomy, Michigan State University, East Lansing, MI 48824, USA}

\author{P.~Gastis}
\affiliation{Departament of Physics, Central Michigan University, Mt. Pleasant, MI 48859, USA}
\affiliation{Joint Institute for Nuclear Astrophysics: CEE,  Michigan State University, East Lansing, MI 48824,
USA}

\author{B.~Gao}
\affiliation{National Superconducting Cyclotron Laboratory, Michigan State University, East Lansing, MI 48824,
  USA}
\affiliation{Joint Institute for Nuclear Astrophysics: CEE,  Michigan State University, East Lansing, MI 48824,
USA}

\author{T.~Ginter}
\affiliation{National Superconducting Cyclotron Laboratory, Michigan State University, East Lansing, MI 48824,
  USA}

\author{C.~J.~Guess}
\affiliation{Departament of Physics and Astronomy, Swarthmore College, Swarthmore, PA 19081, USA}

\author{S.~Lipschutz}
\affiliation{National Superconducting Cyclotron Laboratory, Michigan State University, East Lansing, MI 48824,
  USA}
\affiliation{Joint Institute for Nuclear Astrophysics: CEE,  Michigan State University, East Lansing, MI 48824,
USA}
\affiliation{Departament of Physics and Astronomy, Michigan State University, East Lansing, MI 48824, USA}

\author{B.~Longfellow}
\affiliation{National Superconducting Cyclotron Laboratory, Michigan State University, East Lansing, MI 48824,
  USA}
\affiliation{Departament of Physics and Astronomy, Michigan State University, East Lansing, MI 48824, USA}

\author{A.O.~Macchiavelli}
\affiliation{Lawrence Berkeley National Laboratory, Berkeley, CA 94720, USA}

\author{K.~Miki}
\affiliation{Departament of Physics, Tohoku University, Sendai, Miyagi 980-8578, Japan}


\author{E.~Ney}
\affiliation{Department of Physics and Astronomy, The University of North Carolina at Chapel Hill, Chapel Hill, NC 27599, USA}

\author{S.~Noji}
\affiliation{National Superconducting Cyclotron Laboratory, Michigan State University, East Lansing, MI 48824,
  USA}
\affiliation{Joint Institute for Nuclear Astrophysics: CEE,  Michigan State University, East Lansing, MI 48824,
USA}

\author{J.~Pereira}
\affiliation{National Superconducting Cyclotron Laboratory, Michigan State University, East Lansing, MI 48824,
  USA}
\affiliation{Joint Institute for Nuclear Astrophysics: CEE,  Michigan State University, East Lansing, MI 48824,
USA}

\author{J.~Schmitt}
\affiliation{National Superconducting Cyclotron Laboratory, Michigan State University, East Lansing, MI 48824,
  USA}
\affiliation{Joint Institute for Nuclear Astrophysics: CEE,  Michigan State University, East Lansing, MI 48824,
USA}
\affiliation{Departament of Physics and Astronomy, Michigan State University, East Lansing, MI 48824, USA}

\author{C.~Sullivan}
\affiliation{National Superconducting Cyclotron Laboratory, Michigan State University, East Lansing, MI 48824,
  USA}
\affiliation{Joint Institute for Nuclear Astrophysics: CEE,  Michigan State University, East Lansing, MI 48824,
USA}
\affiliation{Departament of Physics and Astronomy, Michigan State University, East Lansing, MI 48824, USA}

\author{R.~Titus}
\affiliation{National Superconducting Cyclotron Laboratory, Michigan State University, East Lansing, MI 48824,
  USA}
\affiliation{Joint Institute for Nuclear Astrophysics: CEE,  Michigan State University, East Lansing, MI 48824,
USA}
\affiliation{Departament of Physics and Astronomy, Michigan State University, East Lansing, MI 48824, USA}

\author{D.~Weisshaar}
\affiliation{National Superconducting Cyclotron Laboratory, Michigan State University, East Lansing, MI 48824,
  USA}

\date{\today}

\begin{abstract}
The Gamow-Teller strength distribution from ${}^{88}$Sr was extracted from a $(t,{}^{3}\text{He}+\gamma)$ experiment at 115 MeV/$u$ to constrain estimates for the electron-capture rates on nuclei around $N=50$, between and including $^{78}$Ni and $^{88}$Sr, which are important for the late evolution of core-collapse supernovae. The observed strength below an excitation energy of 8 MeV was consistent with zero and below 10 MeV amounted to $0.1\pm0.05$. Except for a very-weak transition that could come from the 2.231-MeV $1^{+}$ state, no $\gamma$ lines that could be associated with the decay of known $1^{+}$ states were identified. The derived electron-capture rate from the measured strength distribution is more than an order of magnitude smaller than rates based on the single-state approximation presently used in astrophysical simulations for most nuclei near $N=50$. Rates based on shell-model and quasiparticle random-phase approximation calculations that account for Pauli blocking and core-polarization effects provide better estimates than the single-state approximation, although a relatively strong transition to the first $1^{+}$ state in $^{88}$Rb is not observed in the data. Pauli unblocking effects due to high stellar temperatures could partially counter the low electron-capture rates. The new data serves as a zero-temperature benchmark for constraining models used to estimate such effects.
\end{abstract}
\pacs{21.60.Cs, 23.40.-s, 25.40.Kv, 26.30.Jk}
\maketitle

\emph{Introduction-} Core-collapse supernovae (CCSNe) are among the most energetic explosions observed in the universe. They contribute to nucleosynthesis, stimulate galactic chemical evolution, and are birth places of neutron stars and black holes \cite{Fryer1999,Heger2003, JANKA200738, BUR13}. A very large fraction of the energy released in CCSNe is in the form of neutrinos, but the small fraction of energy released in the form of visible light is important for probing the mechanism of the explosion. In addition, CCSNe are predicted emission sites of gravitational waves \cite{PhysRevD.78.064056, PhysRevD.95.063019}. Consequently, CCSNe are attractive sites for improving our understanding of the universe through multi-messenger studies \cite{Nak16}. The accurate and detailed description of relevant nuclear physics processes is key to understanding the evolution of CCSNe and interpreting the multi-messenger signals \cite{ARCONES20171}.

Nuclear-weak interaction processes, in particular electron captures (EC), are essential ingredients for simulating and understanding the dynamical evolution of the CCSNe \cite{BETHE1979487, RevModPhys.75.819, Suzuki_2016}.  EC reactions,  on nuclei in the upper $pf$ and $pfg/sdg$-shells are particularly important during the collapse phase \cite{PhysRevLett.91.201102}. It was recently shown \cite{0004-637X-816-1-44,PhysRevC.95.025809, 0954-3899-45-1-014004} that ECs on a group of about 75 nuclei around neutron number $N=50$ between and including  ${}^{78}\text{Ni}$ and ${}^{88}\text{Sr}$ (hereafter referred to as the high-sensitivity region) are responsible for about 50\% of the uncertainties in characteristic parameters such as lepton fraction, entropy, mass enclosed at core bounce, and in-fall velocity \cite{0954-3899-45-1-014004}. Also, the EC rates on nuclei in this mass region could have a significant impact on the nucleosynthesis of trans-iron elements produced in thermonuclear supernovae \cite{refId0}.

EC rates are derived from Gamow-Teller (GT) transition-strength [$B(\text{GT})$] distributions in the $\beta^+$ direction. The EC rates presently used for the nuclei in the high-sensitivity region rely on an approximation that uses a single GT transition for which the strength and excitation energy were fitted to best reproduce EC rates for nuclei in the $pf$ shell near stability \cite{PhysRevLett.90.241102,PhysRevC.95.025805}. This single-state approximation does not account for strong Pauli-blocking effects for heavier nuclei near $N=50$, even for nuclei that are close to stability, and that could strongly reduce the EC rates for neutron-rich nuclei in the high-sensitive region \cite{0004-637X-816-1-44,0954-3899-45-1-014004}.
It is important to verify such effects experimentally and provide data to benchmark and guide theoretical calculations that are used to estimate the EC rates for the astrophysical simulations. At high stellar temperatures, Pauli unblocking are expected to become significant \cite{PhysRevLett.90.241102, PhysRevC.63.032801, PhysRevC.81.015804} and it is important that models used to estimate such effects are first validated at $T=0$. However, for nuclei in which the Gamow-Teller transitions are not completely Pauli blocked, such as for $^{88}$Sr, such temperature-dependent rate effects are expected to be relatively small \cite{PhysRevC.63.032801}.

The only way to experimentally extract GT strength distributions in the $\beta^{+}$ direction for neutron-rich nuclei is through the use of ($n$,$p$)-type charge-exchange (CE) reactions, as the $\beta^{+}$/EC-decay $Q$-values for these nuclei are negative. CE experiments at intermediate beam energies ($\gtrsim 100$~MeV/u) provide an indirect method to extract the $B(\text{GT})$ distributions without $Q$-value constraints, based on a well-established proportionality between the CE cross section at zero momentum transfer and  $B(\text{GT})$ \cite{TADDEUCCI1987125, PhysRevLett.99.202501, PhysRevC.83.054614}. In this letter, we present results of a ($t$,$^{3}$He+$\gamma$) experiment aimed at extracting the Gamow-Teller (GT) transition strength [$B(\text{GT})$; associated with the transfer of $\Delta S=1$ (spin), $\Delta T=1$ (isospin), and $\Delta L=0$ (angular momentum)] from the $N=50, Z=38$ nucleus $^{88}$Sr, which is amongst the most proton-rich nuclei in the high-sensitive region. By combining the $(t,{}^3\text{He})$ CE reaction with high-resolution $\gamma$-ray detection, even weakly excited low-lying GT transitions that may relatively strongly impact the EC rates can be identified, achieving a sensitivity to states with GT strengths of as small as 0.01  \cite{PhysRevLett.112.252501,PhysRevC.92.024312}. This level of experimental sensitivity more or less coincides with the limit on the applicability of the use of charge-exchange reactions for reliably extracting GT strengths. This is due to interference effects between the central $\sigma\tau$ and the tensor-$\tau$ components of the nucleon-nucleon force. For strengths below 0.01, such effects complicate the clean identification of GT transitions from other transitions, and introduce sizeable (30\% for GT strengths of about 0.01) systematic uncertainties \cite{PhysRevC.74.024309,PhysRevC.80.014313,PhysRevLett.112.252501}.

The results discussed here are part of a broader effort to improve the electron-capture rates on nuclei in the high-sensitivity region. These efforts include additional experiments on other nuclei in this region, the incorporation of theoretical nuclear structure models aimed at improving the GT strength distributions used for electron-capture rate calculations, and astrophysical simulations similar to those in Refs. \cite{0004-637X-816-1-44,0954-3899-45-1-014004}.


\emph{Experiment-} A secondary triton beam was produced by fragmentation of a 150~MeV/u ${}^{16}$O primary beam from the NSCL Coupled Cyclotron Facility (CCF) on a  3525-mg/cm$^2$ thick ${}^{9}$Be production target placed at the entrance on the A1900 fragment separator  \cite{MORRISSEY200390}. A 99\%-pure 115~MeV/u triton beam of $4\times 10^6$~pps was generated with a momentum width of 0.5\% (FWHM) by using a 195-mg/cm$^2$ thick Al degrader in the A1900 intermediate image \cite{HITT2006264}. The beam was transported in dispersion-matched mode \cite{SHERRILL1999299,PhysRevC.47.537} to an isotopically enriched ${}^{88}$Sr (99.9\%) foil with a thickness of  19.6~mg/cm$^2$ placed at the S800 Spectrograph \cite{BAZIN2003629} pivot point. Due  to the high reactivity of strontium, a special transfer system was used to insert the target without coming into contact with air. ${}^{3}$He ejectiles produced in the reaction were  momentum analyzed and identified in the S800 focal plane \cite{YURKON1999291}. The particle identification was performed on an event-by-event basis using the energy loss measured in a  5~mm-thick focal-plane scintillator and the time-of-flight relative to the CCF radio-frequency signal. Scattering angles and momenta of the ejectiles at the target location were reconstructed by raytracing the angles and positions measured in two cathode-readout drift chambers (CRDCs) in the S800 focal plane by using a fifth-order  ion-optical inverse matrix calculated in \textsc{cosy infinity} \cite{MAKINO1999338}. Subsequently, the excitation energy ($E_{x}$) of the ${}^{88}$Rb particles was determined in a missing-mass calculation up to 25~MeV with a resolution of 500~keV (FWHM), which is due to the intrinsic resolution that can be achieved in ($t$,$^{3}$He) experiments with a secondary triton beam and the difference in energy loss between tritons and $^{3}$He particles in the $^{88}$Sr target.

 Scattering  angles in the center-of-mass (c.m.) frame were measured in the range of $0^\circ < \theta_{\text{c.m.}}< 5.5^\circ$. A total luminosity of $2\times 10^{32}$ cm$^{-2}$ was achieved over 5 days. Data were acquired for the ${}^{12}\text{C}(t,{}^{3}\text{He}){}^{12}\text{B}(1^+, \text{g.s.})$ reaction by using a 2.6-mg/cm$^2$ thick polystyrene (C$_8$H$_8$)$_n$ target. Its well-known cross section \cite{PhysRevC.83.054614}  was used to calibrate a non-intercepting primary-beam current probe that  served as an absolute measure for the triton beam intensity during the $^{88}$Sr runs.\\
The Gamma-Ray Energy Tracking In-beam Nuclear Array (GRETINA) \cite{PASCHALIS201344,WEISSHAAR2017187}, consisting of thirty-two 36-fold segmented high-purity  Ge detectors mounted on a hemisphere and providing about $1\pi$ solid-angle coverage, was positioned around the $^{88}$Sr target. The use of GRETINA allowed for the precise determination of $\gamma$-ray energies with a high photopeak-detection efficiency (${\sim}4$\% at 2~MeV).

\begin{figure}[!ht]
\centering
\includegraphics[width=0.5\textwidth]{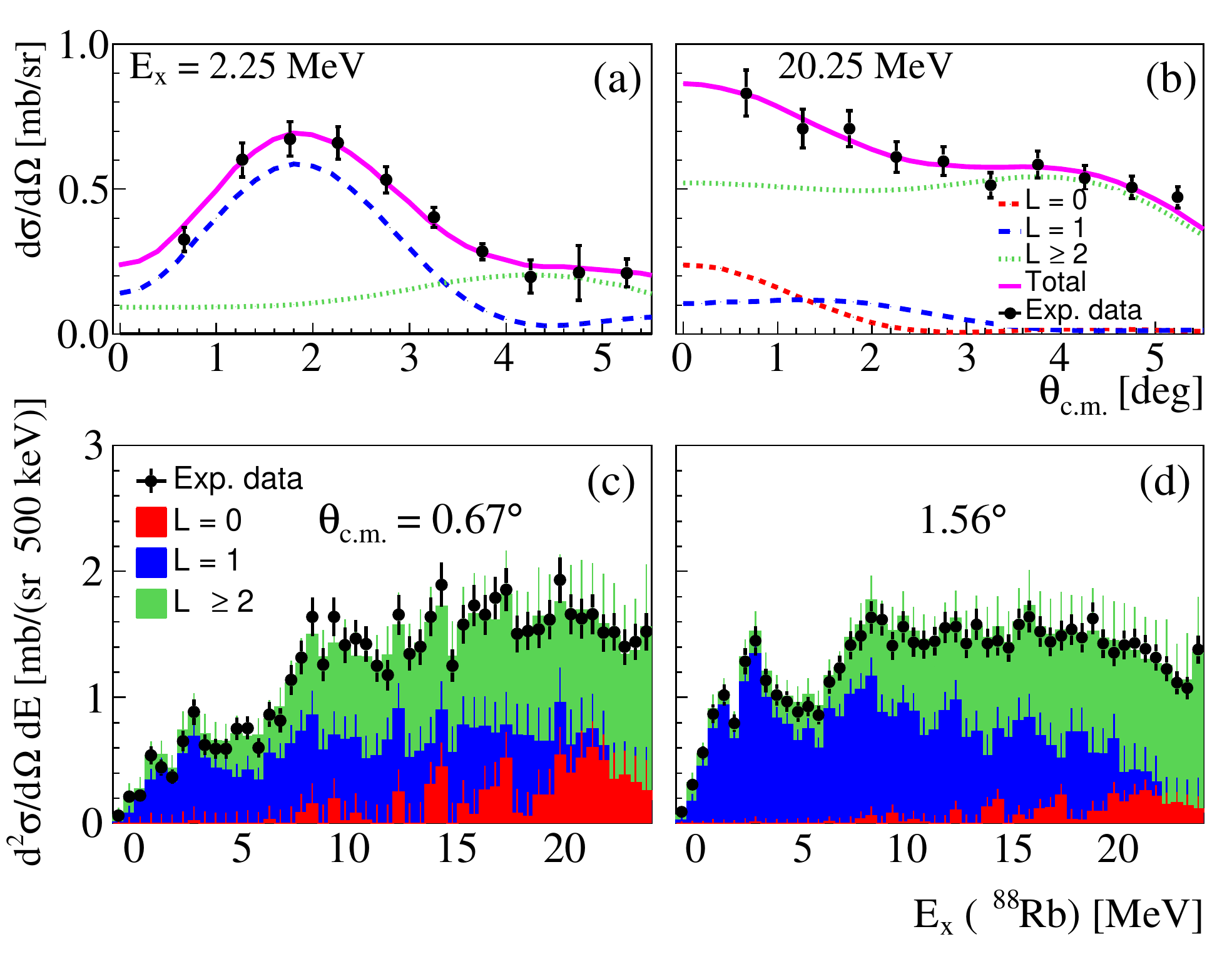}
\caption{\label{mda} (color online) Angular distributions for the ${}^{88}\text{Sr}(t,{}^{3}\text{He})$ reaction at $E_x=2.25$~MeV (a) and 20.25~MeV (b), fitted in the MDA. (Bottom) Double-differential cross sections for  scattering angles in the ranges of $0^\circ < \theta_{\text{c.m.}}<1^\circ$ (c) and $1^\circ < \theta_{\text{c.m.}}<2^\circ$ (d). The colors represent contributions from excitations with different angular momentum transfers.}
\end{figure}

\emph{Results-} Double-differential cross sections for the  ${}^{88}\text{Sr}(t,{}^{3}\text{He})$ reaction were generated in 0.5-MeV wide bins in $E_{x}$. The average statistical error for each bin was 5\%. The systematic error was ${\sim}7$\%, dominated by the uncertainty in the triton beam intensity. Examples for $E_{x}=2.25$ and $20.25$~MeV are shown in Figs.~\ref{mda}(a) and (b), respectively.
To extract the monopole contribution from the cross sections, a multipole decomposition analysis (MDA) \cite{BONIN1984349,ICHIMURA2006446} was performed for each bin in $E_{x}$ by fitting the differential cross section with a linear combination of distorted wave Born approximation (DWBA) angular distributions for angular momentum transfers of $\Delta L=0,1,2$ and 3. The DWBA calculations were performed using the code \texttt{FOLD/DWHI} \cite{fold1, *[{based on }]PETROVICH1977487, *[{modified as described in }]PhysRevC.30.1538, *fold2}. The optical model potential (OMP) parameters  were taken from Ref.~\cite{PhysRevC.67.064612}. Following Ref.~\cite{VANDERWERF1989305}, the depths of the OMP for triton in the incoming channel were scaled by a factor 0.85 from those for ${}^{3}\text{He}$ in the outgoing channel. The effective  nucleon-nucleon interaction of  Franey and Love \cite{PhysRevC.31.488} was double-folded over the transition densities of  $t -{}^{3}\text{He}$  and  ${}^{88}\text{Sr} - {}^{88}\text{Rb}$ systems. The transition densities for $t$ and ${}^{3}$He were taken from  variational Monte Carlo calculations \cite{montecarlo}. For the ${}^{88}\text{Sr} - {}^{88}\text{Rb}$ system, one-body  transition densities  (OBTDs) were generated by using the shell-model code described below. Examples of MDA are shown in Figs.~\ref{mda}(a) and (b). The MDA results for $\theta_{\text{c.m.}}=0.67^\circ$ and $1.56^\circ$ as a function of $E_{x}$ are shown in Figs.~\ref{mda}(c) and (d), respectively. For $E_{x}<8$ MeV, the $\Delta L=0$ contribution of the cross section is consistent with zero within the error bars ($0.07\pm0.1$~mb/sr).  For $E_{x}>10$ MeV, $\Delta L=0$ contributions are observed, but the isovector spin-monopole resonance (IVSMR) is expected to start contributing significantly in this region space \cite{PhysRevLett.108.262503}.

\begin{figure}[!ht]
\centering
\includegraphics[width=0.5\textwidth]{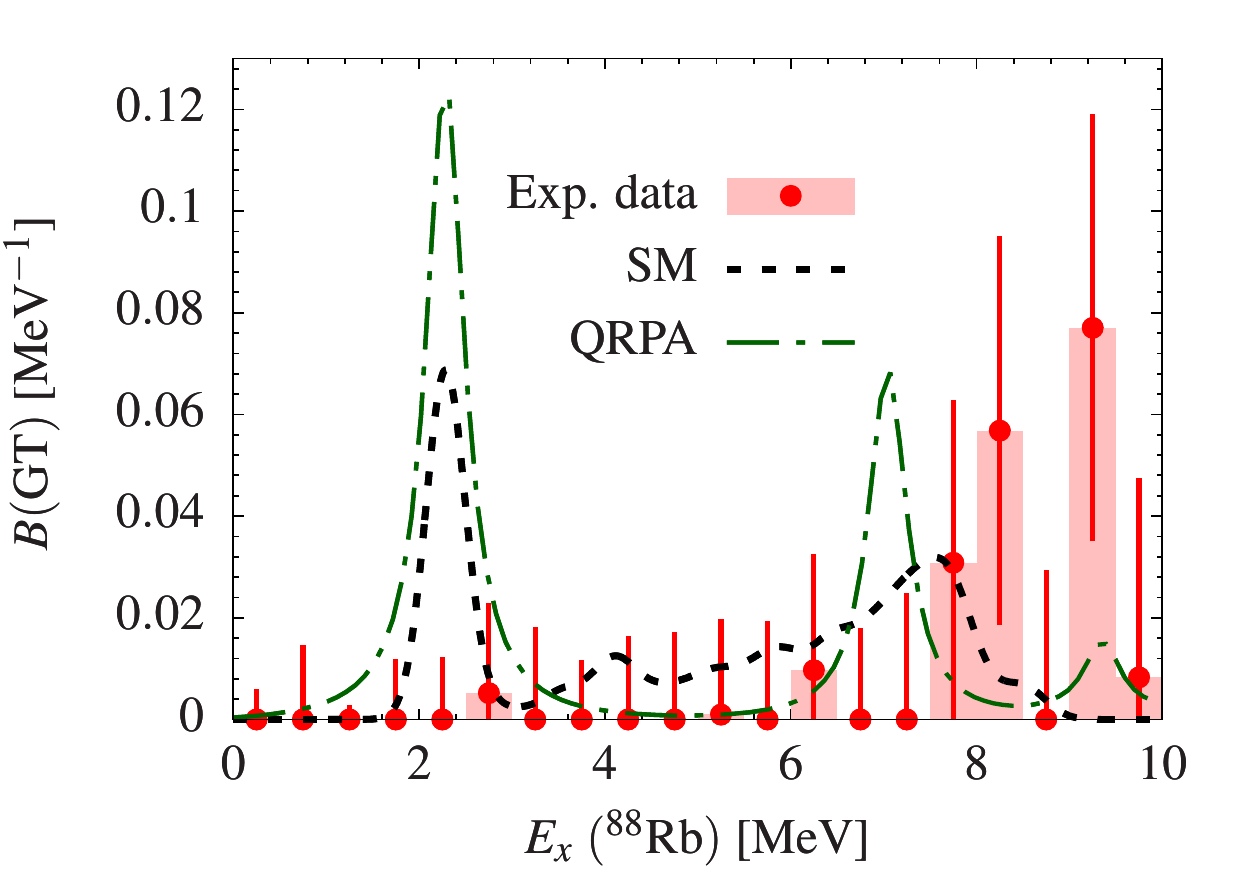}
\caption{\label{bgt} (color online). $B(\text{GT})$ distribution extracted from MDA for $E_{x}<10$~MeV. The error bars denote only the statistical uncertainties. The dashed lines correspond to theoretical calculations, shell model (SM) and QRPA, as described in the text. }
\end{figure}

The $B(\text{GT})$ strength was extracted from the $\Delta L=0$  cross section at  $\theta_{\text{c.m.}}=0^\circ$ by using the proportionality relation: $\sigma_{L=0}(0^\circ) = \hat{\sigma}_{\text{GT}} F(q,\omega) B(\text{GT})$  \cite{TADDEUCCI1987125, PhysRevLett.99.202501, PhysRevC.83.054614}. $\hat{\sigma}_{\text{GT}}$ is the GT unit cross section, which was calculated (5.94~mb/sr) by the mass-dependent empirical relationship of  Ref.~\cite{PhysRevLett.99.202501}, which has an uncertainty of 10\%. $F(q,\omega)$ is a kinematic correction factor that depends on the momentum ($q$) and energy ($\omega$) transfers, and is obtained from DWBA calculations \cite{TADDEUCCI1987125}. Its value was 1.2 (2.1) at $E_{x}=0$ (10) MeV. Fig.~\ref{bgt} shows the extracted  $B(\text{GT})$ distribution in the energy range from 0 to 10~MeV. The only energy bin below the neutron separation energy with non-zero $B(\text{GT})$ is located at $2.75\pm 0.25$~MeV, which correlates with the locations of several $1^{+}$ states in $^{88}$Rb known from the $\beta^{-}$ decay of $^{88}$Kr \cite{PhysRevC.13.1577,MCCUTCHAN2014135}. The summed $B(\text{GT})$ below $E_{x}=10$ MeV is $0.10\pm0.05$, although most of that comes from the region above 8 MeV. This result is significantly lower than the summed $B(\text{GT})$ of $0.7\pm0.1$(stat.)$\pm0.1$(sys.) measured for the $^{90}$Zr($n$,$p$) reaction up to $E_{x}=10$ MeV \cite{YAKO2005193}. Based on transfer reaction experiments \cite{PFEIFFER1986381}, the proton $0g_{9/2}$ occupation is 0.7 (1.0) for $^{88}$Sr ($^{90}$Zr). Therefore, the decrease in GT strength observed below $E_{x}=10$ MeV for $^{88}$Sr as compared to $^{90}$Zr is stronger than expected based on the proton $0g_{9/2}$ occupation number only.

\begin{figure}[!ht]
\centering
\includegraphics[width=0.5\textwidth]{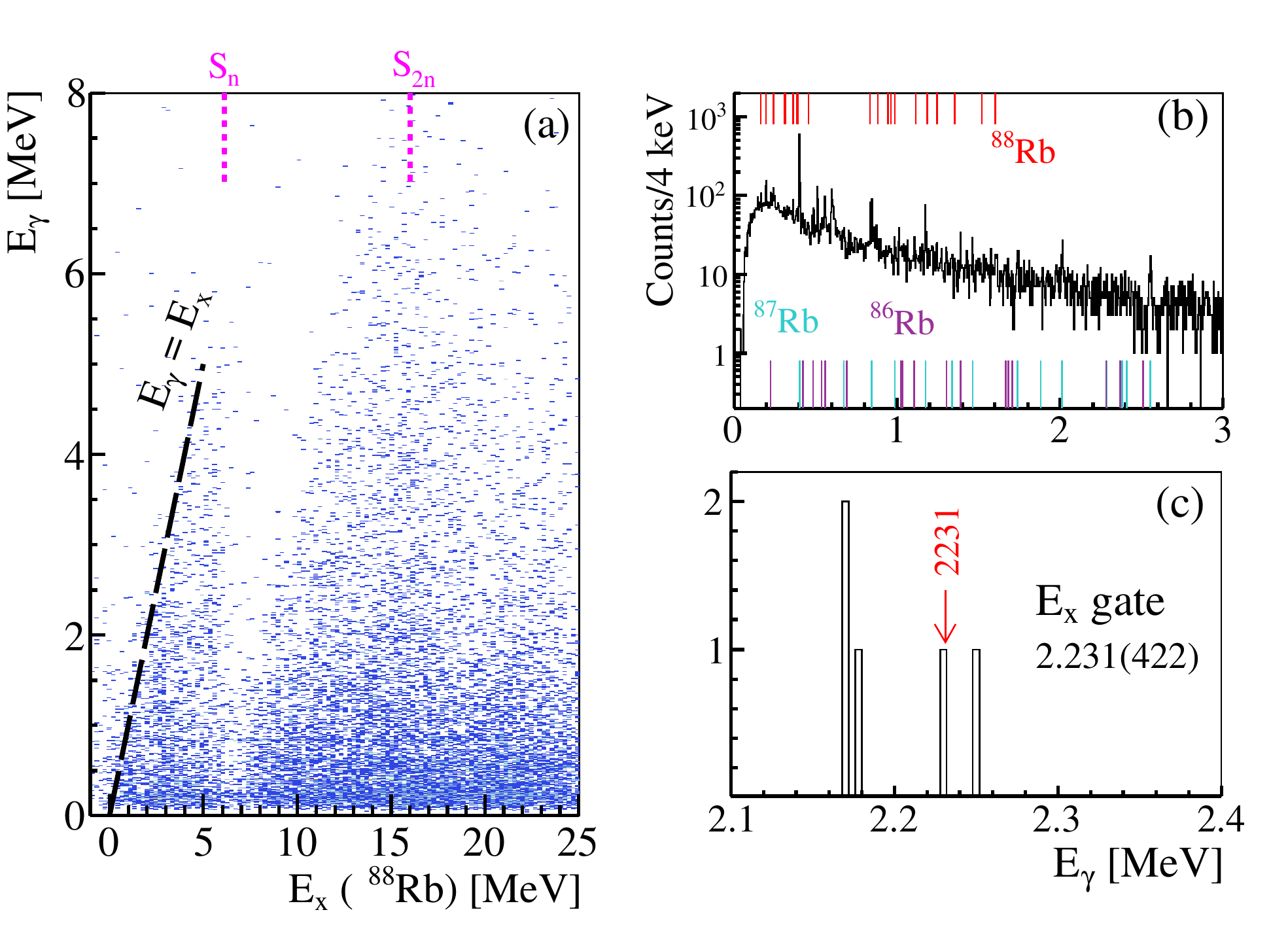}
\caption{\label{corrgp} (color online). (a) Two-dimensional histogram of $\gamma$-ray energy ($E_\gamma$) versus excitation energy ($E_x$) of ${}^{88}$Rb. One and two-neutron separation energies are indicated on top.  (b) Projection on to the $E_\gamma$ axis for $E_x\leq 25$~MeV. The color lines on top and bottom of the figure represent the position of the observed transitions for ${}^{88,87,86}$Rb. (c) $E_\gamma$ spectrum gated at $E_x = 2.231\pm 0.422$~MeV. }
\end{figure}

Additional constraints on the GT strength can be obtained from the ($t$,$^{3}$He$+\gamma$) coincident data. Fig.~\ref{corrgp}(a) shows the two-dimensional histogram that  correlates the energy of $\gamma$ rays ($E_\gamma$) with $E_x$($^{88}$Rb). Due to the wide $E_x$ range covered, $\gamma$-ray transitions from  states  in ${}^{88}$Rb,  ${}^{87}$Rb and ${}^{86}$Rb were observed, as shown in Fig.~\ref{corrgp}(b). The non observation of $\gamma$-rays from ${}^{87}$Kr indicates that the probability of decay by proton emission from ${}^{88}$Rb was very small.

By setting narrow gates on $E_x$ determined from the ($t$,$^{3}$He) reaction, the $\gamma$ spectrum for low $E_x$($^{88}$Rb) was investigated for evidence for the decay from known $1^{+}$ states or for unknown $\gamma$ lines that could stem from previously unknown $1^{+}$ states. No significant signals were found, with the exception of the observation of a single event that could be due to decay from the known 2.231-MeV $1^{+}$ state, as shown in Fig.~\ref{corrgp}(c). This spectrum was obtained by setting a gate on $E_x=2.231\pm0.422$~MeV in the $^{88}$Sr($t$,$^{3}$He) spectrum, where the width of the gate corresponds to $2\sigma$ of the energy resolution. By using a Bayesian analysis \cite{ghosh2007introduction} it was determined that, with an 86\% probability, the credible interval for $B(\text{GT})$ for the 2.231-MeV state ranges from 0 to 0.022, which includes the possibility that the observed count is not due to the decay from this state. The extracted Gamow-Teller strength from the MDA analysis in the relevant excitation energy bin for this transition is $0.006^{+0.02}_{-0.006}$.

The experimental results were compared to shell-model (SM) and quasi-particle random-phase approximation (QRPA) calculations. The shell-model calculations, performed with the code \texttt{NUSHELLX} \cite{nushellx}, assumed a ${}^{78}$Ni  core and a valence space of $(0f_{5/2},  1p_{3/2},1p_{1/2},0g_{9/2})$ for protons and  $(0g_{7/2},  1d_{5/2},1d_{3/2},2s_{1/2},0h_{11/2})$ for neutrons. The proton-proton and proton-neutron two-body matrix elements (TBME) were obtained  from the \texttt{jj44pna} effective interaction \cite{PhysRevC.70.044314} and a renormalized G-matrix using the charge-dependent (CD-Bonn) nucleon-nucleon interaction \cite{PhysRevLett.91.162503}, respectively.
The single-particle energies were determined from the observed single-particle states in  ${}^{89}$Sr.  To  account for the model-space truncation,  the result of the calculation was scaled by a factor $\frac{1}{h}$, where $h$ is a hindrance factor that is a product of two factors: $h_\text{high}$ and  $h_{\text{c.p.}}$  \cite{TOWNER1985402}. $h_\text{high}$ is associated with the admixtures of two-particles two-holes  states with unperturbed energies of $2\hbar\omega$ and higher in the oscillator basis. This factor accounts for the  well-known quenching of the GT transition strength \cite{GAA81,GAA85}.  The empirical value for the  $pf$ model space,  $h_\text{high} = 1.81$ \cite{PhysRevC.53.R2602}, was used.  $h_{\text{c.p.}}$ is due to	the core  polarization for the $0g$ orbital. It accounts for the mixing between $0g_{9/2}$ and $0g_{7/2}$ spin-orbit  partners and depends on the proton occupation number in the $0g_{9/2}$  orbital. $h_{\text{c.p.}}$ is largest when the number of $0g_{9/2}$ protons is small \cite{TOWNER1985402}. An occupation number of 0.58 was calculated for the $\pi 0g_{9/2}$ shell in ${}^{88}$Sr  by using the Ji/Wildenthal effective interaction \cite{PhysRevC.37.1256}, which is close to the experimental value of 0.7 \cite{PFEIFFER1986381}. The hindrance due to the core polarization was taken from the results of Towner in  Ref.~\cite{TOWNER1985402} (Table~5). The value for two protons in $0g_{9/2}$ of $h_{\text{c.p.}}=5.0$, obtained from the $\pi+\rho$ interaction (the range for the three interactions given is 3.5 to 5.9), was used in our calculation. The $0g_{9/2}$ proton number dependence of the hindrance factor $h_\text{high} \times h_{\text{c.p.}}$ leads to a $Z$-dependent hindrance factor that is consistent with that deduced from the $\beta^+$ decay of nuclei with $N=50$ ranging from  ${}^{94}$Ru up to  ${}^{100}$Sn \cite{Ru94,Rykaczewski1985, Plochocki1992, Sn100}.

The QRPA calculation was performed by using the axially-deformed Skyrme finite amplitude method \cite{Avogadro2011,Mustonen2014}. This method has recently been extended to odd-$A$ nuclei in the equal-filling approximation \cite{Shafer2016} and is, therefore, a candidate for calculating GT strengths and EC rates for a large number of nuclei and replacing the EC rates based on the single-state approximation discussed above. The Skyrme functional and single-particle space model are the same as those used in the global calculation of Ref.~\cite{Mustonen2016}, which fixed a single set of parameters including an effective axial-vector coupling constant $g_{A}$ of 1.0.

The theoretical calculations shown in Fig.~\ref{bgt} have been folded with the experimental resolution and the excitation energy of the first $1^{+}$ state was matched to that of the first known  $1^{+}$ state in $^{88}$Rb (at $E_x = 2.231$~MeV). The SM and QRPA calculations both predict a strong transition to the first $1^{+}$ state that is not observed experimentally. The summed strength up to $E_{x}=10$~MeV is 0.12~(0.14) for the SM~(QRPA) calculations. These summed values are consistent with the present data of $0.1\pm0.05$. The results indicate that, besides Pauli blocking, structural effects (core polarization) play an important role in the reduction of the GT strength.

\begin{figure}[!ht]
\centering
\includegraphics[width=0.5\textwidth]{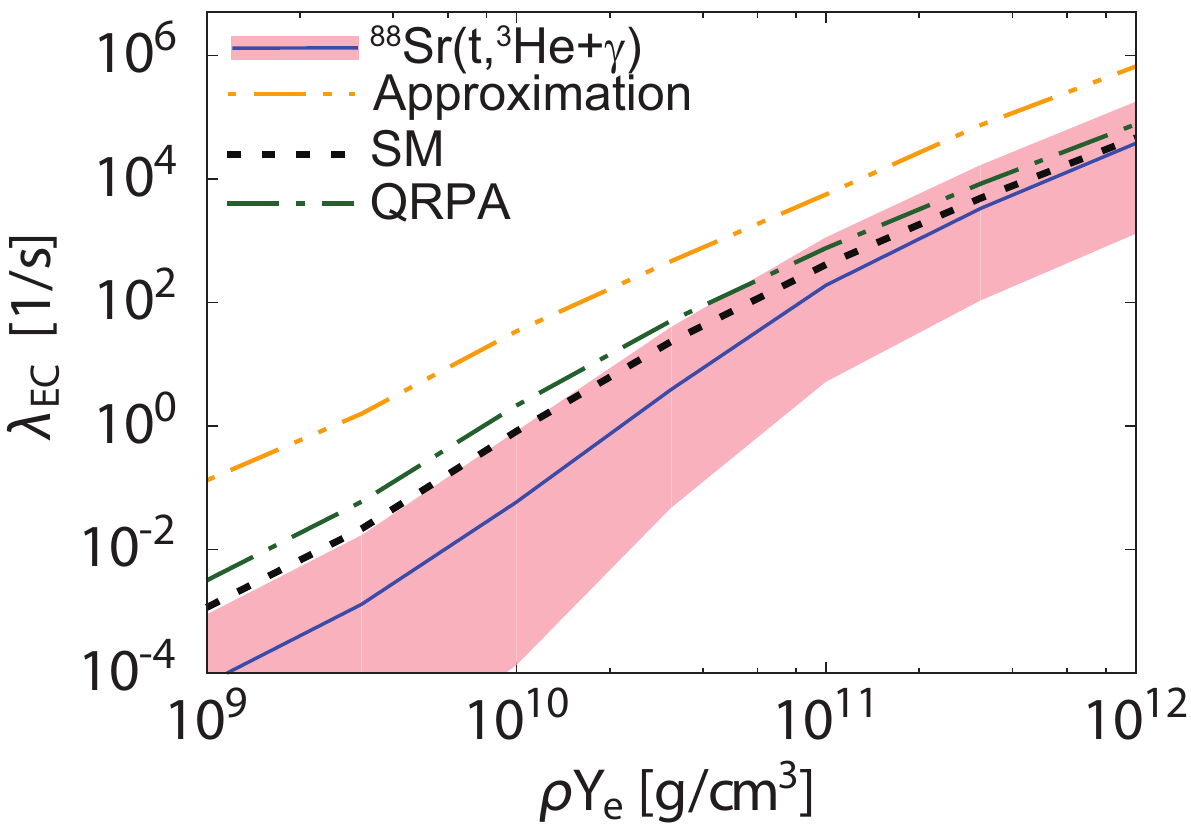}
\caption{\label{ecrates} (color online). EC rates on ${}^{88}$Sr as function of stellar density at a temperature of $10^{10}$~K. The shaded band with solid central curve represents the result based on the $^{88}$Sr($t$,$^{3}$He$+\gamma$) data. The dashed and dot-dashed curves are based on the SM and QRPA calculations, respectively. The dot-dot-dashed line represents the approximate method for estimating the EC rate.}
\end{figure}

Stellar EC rates ($\lambda_\text{EC}$) were calculated based on the formalism in Refs.~\cite{1980ApJSF,  1982ApJF, 1982ApJSF2, 1985ApJF}, in a code previously used in Refs.~\cite{0004-637X-662-2-1188, PhysRevC.86.015809, PhysRevLett.112.252501,PhysRevC.92.024312}. Only transitions from the ground state of ${}^{88}$Sr were considered here. Fig.~\ref{ecrates} shows the calculated EC rates (based on the experimental and theoretical GT strength distributions) during the late stages of CCSN, just prior to the bounce, during which the stellar density ranges from $10^9$-$10^{12}$~g/cm$^3$ and the temperature is ${\sim}10^{10}$~K. Because no known $1^{+}$ state exists below $E_{x}=2$ MeV and the MDA analysis also found no indication for any GT strength up to that energy, the first transition assumed to contribute to the EC rate based on the data was the 2.231~MeV state, with an upperlimit to the strength based on the $\gamma$-decay analysis [$B_{\text{up}}(\text{GT})=0.022$]. The $Q$-value for EC on $^{88}$Sr is $-4.8$~MeV, which means that only at a density of $10^{11}$~g/cm$^{3}$, the Fermi energy of ${\sim}15$~MeV is sufficiently high to cover the strength distribution up to $E_{x}=10$~MeV and that details of the GT strength distribution below that $E_x$ matter up to that density. The higher the density, the less sensitive the EC rate becomes to details of the strength distribution, as most of the strength distribution is below the Fermi energy. Therefore, due to the presence of the relatively strong transition to the first $1^{+}$ state, the EC rates based on the QRPA and SM calculations are higher than the upper limit set by the data for stellar densities below ${\sim}10^{10}$~g/cm$^{3}$. At higher densities, the rates based on the SM and QRPA calculations are within the upper limit set by the data, since the summed strengths up to 10 MeV are, within error bars, consistent. The EC rates calculated based on the single-state approximation (with $B(\text{GT})=4.6$ and a temperature and density dependent effective excitation energy based on Ref.~\cite{PhysRevC.95.025805})  are more than an order of magnitude too high. Considering that ${}^{88}$Sr is the amongst the most proton-rich $N=50$ nucleus in the high-sensitivity region (with the least Pauli blocking), it is very likely that the rates based on the approximation will be also be much too high for the other nuclei in the high sensitivity region.  This has a strong impact on the dynamical evolution during the collapse phase \cite{0004-637X-816-1-44,0954-3899-45-1-014004}.   The drop in lepton fraction during the collapse reduces by 10\% and the enclosed mass at core bounce increases by 10\% when the EC rates for nuclei in the high-sensitivity region are reduced by a factor 10. Finally, we note that although Pauli-unblocking effects due to the high temperature in the collapsing star should increase the EC rates compared to rates shown in Fig.~\ref{ecrates}, the structural (core-polarization) effects are equally important, especially for nuclei such as $^{88}$Sr in which Pauli blocking is not complete at $T=0$, and must be considered in theoretical models used for estimating EC rates at elevated temperatures. Therefore, the present data provides an important zero-temperature benchmark for such theoretical estimates.

\emph{Summary-} The GT transition strength in ${}^{88}$Sr was measured in a high resolution $(t,{}^{3}\text{He}+\gamma)$ experiment to gain insight in EC rates of nuclei near $N=50$ above $^{78}$Ni that are most important during the collapse phase of massive stars prior to the supernova explosion. The extracted  $B(\text{GT})$ is consistent with zero in the energy range from 0 up to $E_{x}=8$~MeV and sums to $0.1\pm0.05$ up to $E_{x}=10$~MeV. SM and QRPA calculations are consistent with this summed strength, but predict a relatively strong transition to a low-lying state not observed in the experiment. As the most proton-rich $N=50$ nucleus in the high-sensitivity region, these results indicate that the EC rates based on a single-state approximation that is used in astrophysical simulations are too high. Although Pauli-unblocking effects due to the high stellar temperatures during the collapse phase partially counter the lowering of the EC rates, the results show that structural effects must be carefully considered as they significantly lower the GT strengths and EC rates. Hence, the present data also serves as a zero-temperature benchmark for theoretical models that can be used to estimate temperature-dependent Pauli-unblocking effects.

\emph{Acknowledgments-} We thank the NSCL staff for their support. This work was supported by the US National Science Foundation (NSF) under Cooperative Agreement PHY-156554 (NSCL), PHY-1430152 (JINA Center for the Evolution of the Elements), and PHY-1811855. GRETINA was funded by the US Department of Energy, in the Office of Nuclear Physics of the Office of Science. Operation of the array at NSCL was supported by DOE under Grants No. DE-SC0014537 (NSCL) and No. DE-AC02-05CH11231 (LBNL). The $^{88}$Sr foil used in this research were supplied by the US Department of Energy Office of Science by the Isotope Program in the Office of Nuclear Physics.

\bibliography{bibliography}  

\end{document}